\documentclass[aps,letterpaper,twocolumn,pra,footinbib,floatfix,showpacs]{revtex4}
\usepackage{amsmath}
\usepackage{amsfonts}
\usepackage{amssymb}
\usepackage[scanall]{psfrag}
\usepackage{psfrag}
\usepackage{graphicx}
\usepackage{color}

\begin{document}
\newcommand{\of}[1]{\left( #1 \right)}
\newcommand{\sqof}[1]{\left[ #1 \right]}
\newcommand{\abs}[1]{\left| #1 \right|}
\newcommand{\avg}[1]{\left< #1 \right>}
\newcommand{\cuof}[1]{\left \{ #1 \right \} }
\newcommand{\bra}[1]{\left < #1 \right | }
\newcommand{\ket}[1]{\left | #1 \right > }
\newcommand{\pil}{\frac{\pi}{L}}
\newcommand{\bx}{\mathbf{x}}
\newcommand{\by}{\mathbf{y}}
\newcommand{\bk}{\mathbf{k}}
\newcommand{\bp}{\mathbf{p}}
\newcommand{\bl}{\mathbf{l}}
\newcommand{\bq}{\mathbf{q}}
\newcommand{\bs}{\mathbf{s}}
\newcommand{\psibar}{\overline{\psi}}
\newcommand{\svec}{\overrightarrow{\sigma}}
\newcommand{\dvec}{\overrightarrow{\partial}}
\newcommand{\bA}{\mathbf{A}}
\newcommand{\bdelta}{\mathbf{\delta}}
\newcommand{\bK}{\mathbf{K}}
\newcommand{\bQ}{\mathbf{Q}}
\newcommand{\bG}{\mathbf{G}}
\newcommand{\bw}{\mathbf{w}}
\newcommand{\bL}{\mathbf{L}}
\newcommand{\ohat}{\widehat{O}}
\newcommand{\up}{\uparrow}
\newcommand{\down}{\downarrow}
\newcommand{\MM}{\mathcal{M}}
\author{Eliot Kapit}
\affiliation{Rudolf Peierls Center for Theoretical Physics, Oxford University, 1 Keble Road, Oxford OX1 3NP, United Kingdom}
\email{eliot.kapit@physics.ox.ac.uk}
\title{Quantum Simulation Architecture for Lattice Bosons in Arbitrary, Tunable External Gauge Fields}


\begin{abstract}

I describe a lattice of asymmetrical qubit pairs in arbitrary dimension, with couplings arranged so that the motion of single-qubit excited states mimics the behavior of charged lattice bosons hopping in a magnetic field. I show in particular that one can choose the parameters of the many-body circuit to reach a regime where the complex hopping phase between any two elements can be tuned to any value by simply adjusting the relative phases of two applied oscillating voltage signals. I also propose specific realizations of our model using coupled three junction flux qubits or transmon qubits, in which one can reach the strongly interacting bosonic quantum Hall limit where one will find anyonic excitations. This model could also be studied in trapped ions, and the superconducting circuits could be used for topological quantum computation.

 
\end{abstract}

\maketitle

\section{Introduction}

Fractional quantum Hall effects \cite{laughlinoriginal,girvin,yoshoika} are among the most profound collections of phenomena to emerge in interacting quantum many-body systems. The elementary excitations in these systems do not act like bosons or fermions; rather, they are {\em anyons}, which in some cases can be used for a robust form of quantum computing \cite{kitaev2003,nayaksimon}. All physical examples of fractional quantum Hall effects are in two dimensional (2D) electron gases. Here we propose a method for linking standard qubit designs which will realize a \textit{bosonic} fractional quantum Hall effect. The rich theoretical literature on bosonic fractional quantum Hall effects suggests that there will be a large number of interesting states \cite{cooperwilkin,hafezi,sorensen,cooper,moller,hormozimoller,juliadiaz2,juliadiaz} that could be explored in our system. 
These include `Pfaffians' and their generalizations.  Furthermore, one could anticipate that some important experiments (such as directly braiding quasiparticles) may be simpler in a qubit array than in a GaAs layer surrounded by AlGaAs.

There are several competing approaches to engineering bosonic fractional quantum Hall effects. One proposal uses Raman lasers to simulate the magnetic vector potential in neutral cold atoms \cite{dalibardRMP,bloch}. 
The technical challenges are, however, quite daunting: new cooling methods need to be designed to offset heating from the Raman lasers, and the most natural probes are indirect.
Another scheme is to use lattices of tiny superconducting grains (charge qubits, \cite{choi,stern,fazio,vanderzant,makhlinschoen,clarkewilhelm}) connected through Josephson junctions. 
%
Suitably low temperatures can be reached in a dilution refrigerator, and the system is readily studied using transport measurements. Unfortunately,  random charge noise, which scales linearly with the interaction strength,  would prevent the quantum Hall regime from being reached without significant local tuning of the potentials on hundreds or thousands of lattice sites. Other proposals include superconducting Jaynes-Cummings lattices \cite{nunnenkamp} and ``photon lattices" of coupled optical waveguides \cite{hafezidemler,umucalilarcarusotto,hafezilukin}, each of which have their own advantages and shortcomings.

I here propose a new and promising approach. Consider a circuit of qubits, with a geometry which naturally maps onto a system of charged bosons hopping in a magnetic field. In order to produce complex hopping matrix elements I study a lattice of coupled asymmetrical pairs of qubits, which I label as $A$ or $B$. I choose device parameters so that excitation energy $\omega_A$ of the $A$ qubits is significantly smaller than the excitation energy of the $B$ qubits, and place a $B$ qubit on each link between neighboring $A$ qubits. Further, I couple them to each other through alternating hopping ($\sigma_{A}^{+} \sigma_{B}^{-} + H.C.$, henceforth referred to as a ``$\pm$" coupling) and potential ($\sigma_{A}^{z} \sigma_{B}^{z}$, a ``$zz$" coupling) terms. Coupling qubits through higher energy auxilliary qubits has been considered previously both theoretically \cite{ashhab} and experimentally \cite{niskanen,niskanenharrabi}. I also apply an external oscillating electromagnetic field of frequency $\omega$ to each qubit, with the relative phase of the signal applied to the $B$ qubits shifted relative to that of the $A$ qubits by a locally tunable $\varphi_s$. Since the $B$ qubits are higher energy than the $A$ qubits, they can be integrated out, leading to complex tunneling matrix elements (the amplitude of a process where the states of neighboring qubit pairs are exchanged) between $A$ qubits with phases that can be tuned to any value by adjusting $\varphi_s$.


As I will describe below, a particularly attractive realization of this architecture would be to use three junction ``flux qubits" (FQs) \cite{mooijorlando,orlandomooij,chiorescunakamura,lyakhovbruder,majerpaauw,gracjar,matsuo,kakuyanagi,bourassa,jiangkane}. The flux qubits are mesoscopic superconducting rings interrupted by three Josephson junctions, placed in a magnetic field which is tuned so that nearly 1/2 of a magnetic flux quantum penetrates the ring. The energies of the flux qubits can be tuned by adjusting this magnetic field, or by varying the areas of the Josephson junctions, so that the $B$ qubits are higher energy than the $A$ qubits as outlined above. We then capacitively couple all the flux qubits to an external, oscillating voltage $V_0 \sin \omega t$, and arrange the couplings so that the phases of the voltage applied to the $B$ qubits are shifted relative to the $A$ qubits. The subtle interplay of the oscillating applied voltages with the mix of charge (capacitive) and phase (Josephson) couplings introduces phase shifts which make these hopping matrix elements complex, mimicking the Peierls phases found for charged particles in magnetic fields.

All of the flux qubits in my design are operated in the regime where the Josephson energy $E_{\rm J}$ is large compared to the charging energy $E_{\rm C}$, so charge noise effects are exponentially suppressed. The system is therefore almost completely insensitive to stray low-frequency electric fields. The many-body excitation gap, a key feature of anyon states, can be measured through the single-qubit response to applied oscillating voltages. The large nonlinearities of the flux qubit devices imply that the first excited states experience an effectively infinite on-site repulsion. I note also that our scheme is not intended to function as a dynamical circuit QED architecture or Jaynes-Cummings model (in contrast to the recent work of Koch \textit{et al} \cite{kochhouck,nunnenkampkoch} and others); the device parameters should be chosen so that the external voltages can be treated as purely classical sources, with no dynamical photons present in our system. Further, though I will not discuss them in any detail here, my proposal could also be studied in trapped ions, perhaps using the methods of Korenbilt \textit{et al} \cite{korenbilt} to engineer the anisotropic spin interactions, or through the digital simulation method detailed in Lanyon \textit{et al} \cite{lanyon}. Finally, since originally submitting this article, I have become aware of a similar method in cold atoms \cite{hauke} for engineering artificial gauge fields through time-dependent drive fields that are out of phase from one lattice site to the next.

The remainder of this paper is organized as follows. In section \ref{general}, I write down the basic coupled qubit Hamiltonian, and outline the conditions under which arbitrary external gauge fields can be simulated. In section \ref{qubits}, I describe three junction flux qubits, and how they can be coupled to obtain the arbitrary complex hopping phases derived in section \ref{general}. Following this, I provide a brief discussion of how these phases might be implemented in transmon qubits \cite{kochyu,schreierhouck,houckkoch,reeddicarlo} as well. I also describe a different regime of operation in the superconducting qubits, where different excited states are mixed by the drive field, which has some advantages over the formulation described in the main body of the text. Having derived the complex hopping phases, in section \ref{LLL} I show how the circuits of the two previous sections can be used as building blocks for exotic boson fractional quantum Hall states. Finally, in section \ref{ex}, I show how a simple arrangement of four qubits could experimentally demonstrate a nonzero effective gauge field, and offer concluding remarks.

\section{General Formalism}\label{general}

\begin{figure}
\vspace{0.5in}
\includegraphics[width=3.0in]{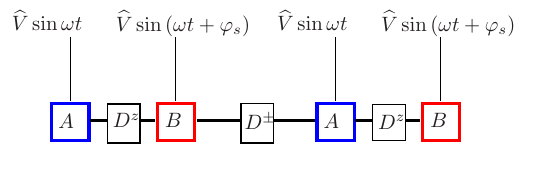}
\caption{(Color online) Basic coupling structure for the $A$ and $B$ qubits. Each site in our many-body lattice would correspond to a single $A$ qubit, which couples to its neighbors through one $B$ qubit per link, joined through alternating hopping ($\pm$) and potential ($zz$) couplings as described in section~\ref{general}. Though drawn in one dimension in the figure, we ultimately intend to construct 2d lattices in this manner, and generalizations to even higher dimensions are also possible.}\label{ABfig}

\end{figure}

\subsection{Berry's Phase of a Rotating Spin}

Before outlining the physics of the larger qubit array, I would first like to discuss a simple example to more straightforwardly elucidate the origin of the complex hopping phases. Specifically, I will consider a pair of spins and examine the Berry's phase effects generated during a process where an excitation is transfered from one spin to its neighbor (whose eigenstates lie on a different axis from the first spin) by rotating both spins about $z$ and then transferred back by rotating both spins about $y$. Let us consider two initially uncoupled spin-$\frac{1}{2}$ degrees of freedom, with the Hamiltonian and eigenstates,
\begin{eqnarray}
H &=& \sigma_{A}^{x} + \cos \theta \sigma_{B}^{z} + \sin \theta \of{\cos \varphi \sigma_{B}^{x} + \sin \varphi \sigma_{B}^{y} }, \\
\ket{0_{A}} &=& \frac{1}{\sqrt{2}} \of{1,-1}, \; \ket{1_{A}} = \frac{1}{\sqrt{2}} \of{1,1}, \nonumber \\
\ket{0_{B}} &=& \frac{1}{\sqrt{2-2 \cos \theta}} \of{e^{i \varphi} \of{-1 + \cos \theta},\sin \theta}, \nonumber \\
\ket{1_{B}} &=& \frac{1}{\sqrt{2+2 \cos \theta}} \of{1+ \cos \theta,e^{-i \varphi} \sin \theta}. \nonumber
\end{eqnarray}
Let us assume that initially spin $A$ is excited and spin $B$ is in its ground state. We first act with the operator $\sigma_{A}^{z} \sigma_{B}^{z}$ to transfer the excitation from $A$ to $B$, and assume energy is conserved in this process so that the final state after acting with $\sigma_{A}^{z} \sigma_{B}^{z}$ is $\ket{0_A 1_B}$; since the $B$ spin is quantized along a different direction from $A$, the excitation must rotate to be transferred to the $B$ spin. We then act with $\sigma_{A}^{y} \sigma_{B}^{y}$ to transfer the excitation back; the resulting matrix element $\MM$ for the entire process is
\begin{eqnarray}
\MM &=& \bra{1_{A} 0_{B}} \sigma_{A}^{y} \sigma_{B}^{y} \ket{0_{A} 1_{B} } \bra{0_{A} 1_{B}} \sigma_{A}^{z} \sigma_{B}^{z} \ket{ 1_{A} 0_{B} } \\
&=& \sin \theta \of{ \cos \varphi + i \sin \varphi \cos \theta }. \nonumber
\end{eqnarray}
For $\theta \neq \pi/2$ and $\varphi \neq 0,\pi$, $\MM$ is complex, and the resulting phase can be understood as a consequence of the Berry's phase acquired by a rotating spin, though we note of course that the Berry's phase discussed here is only an analogy, since we are considering the action of pairs of operators and not continuous, adiabatic changes to the system's wavefunction. When a spin $m$ is rotated along a closed path, the resulting phase is equal to $m$ times the area subtended by the path on the unit sphere. In this case, we have two spins which rotate, but both end in the same states in which they started, so we obtain a gauge-invariant phase equal to the sum of the phases picked up by both spins. The area subtended by $A$ is just $\pi$, but the area subtended by $B$ depends on the projection of $\sigma_{y}$ and $\sigma_{z}$ onto its quantization axis, and thus depends on $\varphi_s$, yielding the result above. Note that if we'd acted with $\sigma_{A}^{z} \sigma_{B}^{z}$ or $\sigma_{A}^{y} \sigma_{B}^{y}$ twice instead of using a combination of the two, the outcome would necessarily be real, since $\MM$ would be the product of a matrix element and its Hermitian conjugate. In the Berry's phase picture, the phase is zero simply because the path of each spin's rotation would be a 1d line, and thus each area is zero. Both inequivalent eigenstates and anisotropic operations are necessary for spin transfer matrix element to be complex. 

It is precisely this effect--the phase picked up by a spin which rotates as it propagates in space--which I will use to engineer artificial hopping phases in our lattice. Specifically, imagine the case in which we had two (identical) $A$ spins with a $B$ spin in between them, and after acting with $\sigma_{A1}^{z} \sigma_{B}^{z}$ to pass an excitation from $A1$ to the $B$ spin, we then act with $\sigma_{A2}^{y} \sigma_{B}^{y}$ to transfer the excitation to the second $A$ qubit instead of sending it back to the first. Since the $A$ spins are identical, the matrix element $\MM$ should be the same as the one derived above, and therefore by letting $B$ spins mediate a hopping coupling, we can introduce tunable phases in a lattice of $A$ spins.

Engineering this structure in a real spin (or qubit) lattice is by no means trivial. For real spins, one could introduce a spatially varying magnetic field to generate the inequivalent local eigenstates, but adding the anisotropic spin-spin interactions ($\sigma_{A}^{z} \sigma_{B}^{z}$ or $\sigma_{A}^{y} \sigma_{B}^{y}$ instead of $\mathbf{S}_{A} \cdot \mathbf{S}_{B}$) is very difficult. Conversely, for a more general lattice of qubits, generating passive anisotropic couplings is often straightforward, but generating inequivalent local eigenstates is not. I here demonstrate that coupling the qubits to a continuously oscillating monochromatic external field can introduce the required rotations, provided that the phases of the signals applied to the $B$ qubits are different from those applied to the $A$ qubits. By adjusting these phases at a local level, we can independently tune the tunneling phase between any linked sites on the lattice, and can thus simulate any desired external gauge field, at least in principle.

\subsection{Qubit Coupling Hamiltonian}

We will consider a lattice of qubits, arranged such that there is a higher energy $B$ qubit between each pair of linked $A$ qubits. We shall assume throughout that the following conditions hold:

(1) The nonlinearities of each physical system which we use as a qubit are large enough that we can consider them to be purely two-level systems, and ignore all eigenstates besides $\ket{0}$ and $\ket{1}$. This requirement ultimately constrains the magnitudes of the couplings between qubits, which must be small compared to the physical devices' absolute nonlinearities. I describe an alternate regime, where states $\ket{2}$ and higher are considered and mixed by the drive fields, later in this work.

(2) The qubits can be coupled to an external electromagnetic field. We shall further require that the electromagnetic field operator $\widehat{V}$ (which could represent the coupling to magnetic fields as well) has no expectation value in either state, so $\bra{0}\widehat{V}\ket{0} = \bra{1}\widehat{V}\ket{1} = 0$. These fields will always be present in the qubit array Hamiltonian, and we will treat them in the standard rotating wave approximation.

(3) We must be able to introduce two types of coupling between the qubits, so that the qubit-qubit Hamiltonian takes the form
\begin{eqnarray}
H_{int} = D^{\pm} \of{ \sigma_{A}^{+} \sigma_{B}^{-} + \sigma_{A}^{-} \sigma_{B}^{+} } + D^{z} \sigma_{A}^{z} \sigma_{B}^{z}.
\end{eqnarray}
We must have independent control over both $D^\pm$ and $D^z$ for our method to succeed. Note that any physical coupling between the qubits will typically include terms which violate number conservation. However, when we transform to the rotating frame when the external oscillating voltage is applied, the terms in $H_{int}$ are unchanged but anomalous terms such as $\sigma_{A}^{-} \sigma_{B}^{z}$ or $\sigma_{A}^{+} \sigma_{B}^{+}$ will become rapidly oscillating and can be dropped from the low-energy Hamiltonian.

(4) We must be able to tune the relative phase $\varphi_s$ of the external electromagnetic field applied to $B$ qubits relative to the $A$ qubits, as shown in fig.~\ref{ABfig}. If $\varphi_s \neq 0,\pi$ then time reversal symmetry is broken, since we cannot chose a zero point for the time $t$ so that both $V_{A} \of{t}= V_{A} \of{-t}$ and $V_{B} \of{t} = V_{B} \of{-t}$. Breaking time reversal symmetry is a basic requirement for obtaining nontrivial effective gauge fields.

These requirements could be fulfilled by a large number of physical systems, including spin qubits, trapped ions, and superconducting devices, which will be the focus of this work. Let us now consider the Hamiltonian of a given qubit pair, $H_{AB}$. Before turning on the oscillating fields, our qubit Hamiltonian is
\begin{eqnarray}
H_{AB}^{0} &=& \frac{\omega_{A}}{2} \sigma_{A}^{z} + \frac{\omega_{B}}{2} \sigma_{B}^{z} \\
& & + \cuof{ D^{\pm} \of{ \sigma_{A}^{+} \sigma_{B}^{-} + \sigma_{A}^{-} \sigma_{B}^{+} } \; {\rm{or}} \; D^{z} \sigma_{A}^{z} \sigma_{B}^{z} }. \nonumber
\end{eqnarray}
We now turn on the oscillating fields. When acting on $A$ or $B$, we have:
\begin{eqnarray}
\widehat{V} =  \Omega_{A/B} \sigma_{A/B}^{y},
\end{eqnarray}
with $ \Omega_{A/B} = \bra{1_{A/B}} \widehat{V} \ket{0_{A/B}}$, which we choose to be real. We now examine
\begin{eqnarray}\label{actV}
\widehat{V} \sin \omega t &=& \frac{\Omega_{A/B}}{2} \of{e^{i \omega t} \sigma_{A/B}^{-} + e^{- i \omega t} \sigma_{A/B}^{+}} \\
& & + \frac{\Omega_{A/B}}{2} \of{e^{-i \omega t} \sigma_{A/B}^{-} + e^{ i \omega t} \sigma_{A/B}^{+}}. \nonumber
\end{eqnarray}
We now transform to the rotating frame by applying the unitary transformation $\ket{\psi} \to \exp -i \frac{\omega}{2} \of{\sigma_{A}^{z} + \sigma_{B}^{z}} t \ket{\psi}$. The time dependence of terms on the first line of (\ref{actV}) is cancelled out, leaving us with $\Omega_{A/B} \sigma_{A/B}^{x}/2$ plus a set of terms which are rapidly oscillating with frequency $2\omega$. We now make the rotating wave approximation (RWA) to neglect these terms. After transforming to the rotating frame and invoking the RWA, $H_{AB}$ is:
\begin{eqnarray}\label{HAB}
H_{AB} &=& \frac{\of{\omega_{A} - \omega}}{2} \sigma_{A}^{z} + \frac{\of{\omega_{B} - \omega}}{2} \sigma_{B}^{z} \\ & & + \frac{\Omega_{A}}{2} \sigma_{A}^{x} + \frac{\Omega_{B}}{2} \of{ \cos \varphi_{s} \sigma_{B}^{x} + \sin \varphi_{s} \sigma_{B}^{y} } \nonumber \\ & & + \cuof{ D^{\pm} \of{ \sigma_{A}^{+} \sigma_{B}^{-} + \sigma_{A}^{-} \sigma_{B}^{+} } \; {\rm{or}} \; D^{z} \sigma_{A}^{z} \sigma_{B}^{z} }. \nonumber
\end{eqnarray}
From now on we will assume $\omega$ is tuned to resonance with the $A$ qubits, so that $\omega = \omega_{A}$ and the single-site Hamiltonian for the $A$ qubits is just $\Omega_{A} \sigma_{A}^{x}/2$.

To construct the full qubit lattice, we wire the qubits as in fig.~\ref{ABfig}, so that the connection between any pair of neighboring $A$ qubits consists of a $zz$ coupling to a $B$ qubit followed by a $\pm$ coupling to the other $A$ qubit. For simplicity, we will ignore cases where $A$ qubits are coupled directly; such couplings will produce either neighbor-neighbor potential interactions or real-valued hopping matrix elements, depending on their structure. We assume that the energy difference $E_{B} - E_{A} =  \sqrt{\of{\omega_{B} - \omega_{A} }^{2} +  \Omega_{B}^{2} } -  \Omega_{A} \equiv \delta E$ is large compared to $D^{\pm}$ and $D^{z}$, so that we can treat the $A-B$ coupling perturbatively. We now eliminate the $B$ qubits using second order perturbation theory; noting that all $A$ qubits are identical, the resulting Hamiltonian, to order $D^{2}/\delta E$, is given by:
\begin{eqnarray}\label{Hlat}
H &=& \sum_{ij} \of{J_{ij} a_{i}^{\dagger} a_{j} + H. C.} +  \tilde{\Omega}_{A} \sum_{i} a_{i}^{\dagger} a_{i}, \\
J_{ij} &=& -\frac{D_{ij}^{z} D_{ij}^{\pm}}{2 \delta E} \sin \theta \of{ \cos \varphi_{s(ij)} + i \cos \theta \sin \varphi_{s(ij)} }, \nonumber \\
\cos \theta &=& \frac{\omega_{B} - \omega_{A}  }{  \sqrt{\of{\omega_{B} - \omega_{A} }^{2} +  \Omega_{B}^{2} } }. \nonumber
\end{eqnarray}
Here $a_{i}^{\dagger}/a_{i}$ creates/annihilates an excitation in the $A$ qubit at site $i$, and $\tilde{\Omega}_{A}$ is equal to $\Omega_{A}$ plus $O \of{J}$ shifts which depend on the coordination number of the lattice and magnitudes of the couplings. Since the qubits are spin-$\frac{1}{2}$, we have an effective hard-core constraint, so $a_{i}^{\dagger} \ket{1_i} = 0$. If we now identify
\begin{eqnarray}\label{Peierls}
{\rm{arg}}J_{ij} \equiv q \int_{r_{i}}^{r_{j}} \mathbf{A} \cdot d\mathbf{r},
\end{eqnarray}
we see that the complex phases of $J$ are identical to the Peierls phases of a charged particle moving on a lattice in an external gauge field $\mathbf{A}$. Further, if we choose parameters so that the $B$ qubits are far off-resonance, $\theta$ will be small and
\begin{eqnarray}\label{Jphase}
J_{ij} \to -\frac{D_{ij}^{z} D_{ij}^{\pm}}{2 \delta E} \theta e^{i \varphi_{s(ij)} } + O \of{\theta^{3}}.
\end{eqnarray}
In this regime, we can freely adjust the phase of $J$ without significantly altering its magnitude, and can thus simulate any time-dependent external gauge field configuration we desire, simply by adjusting the $B$ qubit phase shifts $\varphi_{s(ij)}$ at each link. 

Before continuing, it is worth keeping in mind that the rotating wave approximation is simply the zeroth-order term in a power series in $\Omega_{A/B}/\omega$, and is therefore not an exact description of the system's dynamics. Corrections to the RWA have been treated in many ways, but for our purposes the treatment of Thimmel \textit{et al} \cite{thimmel} is the most useful, since it details the effect of terms beyond the RWA on the time-independent rotating frame effective Hamiltonian. Generalizing their result, we obtain an effective correction term at first order in $\Omega_{A/B}/\omega$
\begin{eqnarray}
\delta H  &=&  \frac{3\Omega_{A}^{2}}{8\omega}  \sigma_{A}^{z} -   \frac{\Omega_{A} \of{\omega_{A}-\omega}}{4\omega} \sigma_{A}^{x} \\
& & + \frac{3\Omega_{B}^{2}}{8\omega}  \sigma_{B}^{z} -   \frac{\Omega_{B} \of{\omega_{B}-\omega}}{4\omega} \of{ \cos \varphi_{s} \sigma_{B}^{x} + \sin \varphi_{s} \sigma_{B}^{y} }. \nonumber
\end{eqnarray}
These small shifts can be eliminated by further tuning of the device parameters and applied frequencies, and should not change the basic physics of the system or its artificial gauge field. In particular, in the flux qubits I will describe below, $\omega$ is 15-30 times larger than the Rabi frequencies $\Omega_{A/B}$, so these corrections are strongly suppressed.

In addition to these corrections, in a physical qubit system the applied voltage $\hat{V}$ will include matrix elements that mix the qubit's basis states with higher excited modes \footnote{In the flux qubits we focus on in this paper the lowest order matrix element of this type is $\bra{2}\hat{V}\ket{0}$; in transmons and other single-junction devices it is $\bra{2}\hat{V}\ket{1}$.}. These transitions can be treated in perturbation theory by integrating out the higher modes, and produce $\sigma^{z}$ terms which scale as $\Omega^{2}/\of{\omega - \omega_{ij}}$, where $\omega_{ij}$ is the energy difference between the coupled states. These corrections can become significant if the Rabi frequencies $\Omega$ approach the absolute nonlinearities of the qubit spectra, but for flux qubits these nonlinearities are large and mixing with higher excited states can be ignored. Alternatively, the higher excited states can be used to our advantage by driving the system near a different resonance to leave the ground state unchanged; I will describe this approach in section \ref{12drive}.

The ability to engineer artificial gauge fields of any desired configuration has tremendous potential to unlock new physics, and I will discuss the most natural application, simulating a uniform magnetic field to realize strongly interacting bosons in the quantum Hall regime, later in the work. Before doing so, however, I will first describe a possible implementation of this architecture in superconducting flux qubits. While flux qubits are certainly not the only-- or even necessarily the best-- qubits to use for this purpose, our proposal will demonstrate that a fairly robust implementation of my architecture can be realized using device parameters from previous experiments. Thus, small lattices should be within reach of current technology.

\begin{widetext}

\begin{figure}
\includegraphics[width=6in]{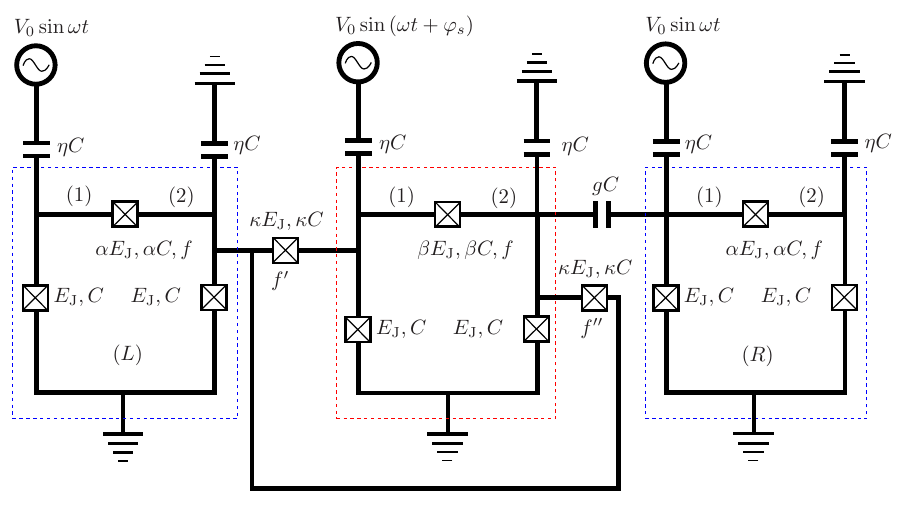}
\caption{(Color online) Basic circuit architecture. The regions enclosed in dashed boxes are three-junction flux qubits, which are connected to a physical ground. The blue ($A$, left and right) and red ($B$, center) qubits differ from each other by a rescaling of the area of the central Josephson junction, which is tuned so that the $B$ qubits have higher energy excitations. A magnetic field penetrates the plane so that $f$ flux quanta are enclosed by each ring. An oscillating voltage $V_{E} \of{t}$ is applied near resonant transitions to both qubits, mixing their ground and first excited states. Excitations in the $A$ flux qubits can tunnel through the $B$ qubits to each other; the oscillating voltage will make this transition matrix element complex. The qubit properties and the couplings between them are discussed in section~\ref{qubits}.}\label{qubitfig}
\end{figure}

\end{widetext}

\section{Qubit Implementations}\label{qubits}

\subsection{Flux Qubits}

The three-junction flux qubit consists of a superconducting ring interrupted by three Josephson junctions as shown in fig.~\ref{qubitfig}, with one junction whose area is rescaled by $\alpha$ relative to the other two. A constant, tunable magnetic flux bias of $f \neq 1/2$ flux quanta is applied through the loop. We choose bottom third of the ring to be ground (which will be a physical ground in our case) with phase $\phi=0$, then the two remaining degrees of freedom of the flux qubit are the phases $\phi_1$ and $\phi_2$ of the other two superconducting regions. The derivation of the flux qubit Hamiltonian is descrbed in detail in Orlando \textit{et al} \cite{orlandomooij}; in terms of the phases $\phi_1$ and $\phi_2$, the flux qubit Hamiltonian $H_{FQ}$ is
\begin{eqnarray}\label{HFQ}
H_{FQ}  &=& \frac{ \of{1+ \alpha + \eta} \of{Q_{1}^{2} + Q_{2}^{2} } + 2 \alpha Q_{1} Q_{2}   }{\of{1+\eta} \of{1 + 2 \alpha + \eta} C } \\ 
& & - E_{\rm J} \sqof{ \cos \phi_1 + \cos \phi_2 + \alpha \cos \of{ 2 \pi f + \phi_1 - \phi_2 } } \nonumber \\
& & + \frac{2 \eta \of{\alpha Q_{2} + \of{1+\alpha + \eta} Q_{1} } V_{0} \sin \omega t }{\of{1+\eta} \of{1 + 2 \alpha + \eta}}. \nonumber 
\end{eqnarray}
Here, $Q_{j} = - 2 e i \partial/ \partial \phi_j$, $E_{\rm J}$ is the Josephson energy of the Josephson junctions and $f$ is the total magnetic flux through the loop in units of the magnetic flux quantum $\Phi_0$. The terms on the third line of (\ref{HFQ}) represent the coupling of the flux qubit to the applied voltage $V_0 \sin \omega t$. For the moment, we will consider this Hamiltonian with $V_0 = 0$.

We let $\phi_\pm = \of{\phi_1 \pm \phi_2 }/2$. For $f \neq 0$, the symmetry between $\phi_1$ and $\phi_2$ is broken, and for $f$ close to 1/2, the ground and first excited states are distinguished by their behavior along the $\phi_-$ direction, as excitations along $\phi_+$ are significantly more expensive. The typical excitation energy for $0.4 < \alpha < 0.6$ and $0.5 < f < 0.55$ is $\omega_{FQ}/2\pi = 12 - 30 \rm{GHz}$ for $E_{\rm J}/h \sim 200 \rm{GHz}$ and $E_{\rm C} =e^2/2C = E_{\rm J}/40$, and the nonlinearities of the spectrum are all reasonably large. In this work we will only consider flux qubits operated at the symmetry point of $f=1/2$, in which case the ground and first excited states are both even along $\phi_+$ and even or odd, respectively, along $\phi_-$. From this, we can readily translate operators in the phase basis to Pauli matrices acting in the qubit basis. We will define the following compact notation for matrix elements:
\begin{eqnarray}\label{Mdef}
\MM_{\widehat{O},s}^{ij} \equiv \bra{i_{s}} \widehat{O} \ket{j_{s}} \; \rm{e.g.} \; \MM_{Q_{1},A}^{01} = \bra{0_A} Q_{1} \ket{1_A}. 
\end{eqnarray}
In this notation, we have:
\begin{eqnarray}\label{relations}
Q_{j} &\to & 2 e \of{-1}^{j} \MM_{\partial_{\phi_{-}}}^{01} \sigma^{y}, \\
\sin \phi_{j} &\to & \of{-1}^{j} \MM_{\sin \phi_{1}}^{01} \sigma^{x},  \nonumber \\
\cos \phi_{j} & \to & \frac{\MM_{\cos \phi_1}^{11}+\MM_{\cos \phi_1}^{00}}{2} \mathbf{1} + \frac{\MM_{\cos \phi_1}^{11}-\MM_{\cos \phi_1}^{00}}{2} \sigma^{z}. \nonumber
\end{eqnarray}
For consistency, all matrix elements $\MM$ are calculated between the $V_0=0$ (non-rotating) eigenstates of the flux qubit Hamiltonians.

Let us now turn to the coupling Hamiltonian between the qubits shown in fig.~\ref{qubitfig}. Direct and indirect coupling of flux qubits, including through intermediary qubits \cite{niskanen}, has been considered theoretically and demonstrated experimentally \cite{gracjar,izmalkov,majerpaauw,niskanenharrabi,ashhab,harrisberkley,paauwfedorov,harrisjohnson}. We label the two $A$ qubits by $L$ and $R$. The coupling of the $B$ qubit to the right qubit is a simple capacitive coupling, and so is given by a constant times $\sigma_{B}^{y} \sigma_{R}^{y}$, which becomes a $\pm$ coupling in the rotating frame:
\begin{eqnarray}\label{capterm}
H_{BR} = \frac{8 E_{\rm C} \MM_{\partial_{\phi_{1}}, B}^{01} \MM_{\partial_{\phi_{1}}, R}^{10}}{ \of{1+2\alpha + \eta} \of{1+2\beta + \eta}} \of{\sigma_{B}^{+} \sigma_{R}^{-} + \sigma_{B}^{-} \sigma_{R}^{+}}.
\end{eqnarray}
It is important to note that both $\sigma^{x} \sigma^{x}$ and $\sigma^{y} \sigma^{y}$ become $\pm$ couplings in the rotating frame, as the components of them which lead to net creation or destruction of excitations are rapidly oscillating and should be dropped. The coupling between the left qubit and the $B$ qubit consists of two Josephson junctions; since these junctions define closed loops through ground, they pick up flux biases $f'$ and $f''$ from the external magnetic field. For simplicity, we choose the wiring geometry so that these biases are both zero mod $2\pi$. When we write the coupling between $L$ and $B$ as a set of Pauli matrices, the $\pm$ terms vanish due to the sign flips in (\ref{relations}), but the $zz$ term survives:
\begin{eqnarray}
H_{LB} &=& -2 \kappa E_{J} \of{\MM_{\cos \phi_1,L}^{11}-\MM_{\cos \phi_1,L}^{00}} \\ & & \times \of{\MM_{\cos \phi_1,B}^{11}-\MM_{\cos \phi_1,B}^{00}} \sigma_{L}^{z} \sigma_{B}^{z}. \nonumber
\end{eqnarray}
Alternately, one could obtain a pure $zz$ coupling by simply placing a single Josephson junction between a pair of regions, and choosing the wiring geometry so that the flux bias $f'$ is nonzero, leading to an interaction term of the form $- \kappa E_{J} \cos \of{\phi_{L2} - \phi_{B1} + 2\pi f'}$ plus a capacitive term with the same structure as (\ref{capterm}). One could then tune $f'$ so that the $\pm$ components of the $xx$ and $yy$ terms from these couplings interfere with each other, leaving only the $zz$ part of the coupling.

\begin{figure}
\includegraphics[width=3in]{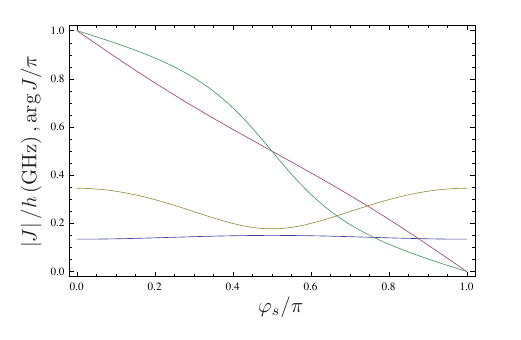}
\caption{(Color online) Magnitude and phase of $J$ for the device parameters given in eq. (\ref{Jpars}) at the resonance point $\omega = \omega_{A}$. The blue and purple curves are $\abs{J}$ and $\arg{J}/\pi$, respectively, for $V_{0} = 0.25 \rm{mV}$; the yellow and green curves are the same quantities for $V_{0} = 0.5 \rm{mV}$. $J$ can be made significantly larger by increasing $\alpha$ and working away from the $f=1/2$ symmetry point, but the physical device nonlinearities are smaller in that regime and the system becomes more susceptible to fluctuations in the external magnetic field.}\label{Jvsphifig}
\end{figure}

We are now in a position to plug in numbers and evaluate $J$ for this architecture. Consider flux qubits wired as in fig.~\ref{qubitfig}. If we choose the realistic device parameters listed below, taking into account the single-qubit energy shifts from the $D^{z}$ coupling gives us:
\begin{eqnarray}\label{Jpars}
I_{c} &=& 400 {\rm{nA}}, \; C = 3.25 {\rm{fF}}, \; \alpha = 0.5, \; f=0.5, \\
\eta &=& 0.1, \; \beta = 0.45, \; \kappa = 0.2, \; g =  0.2 \nonumber  \\
E_{\rm J}/h & = & 200 {\rm{GHz}} = 33 E_{\rm C}/h, \; \omega_{A} = 2\pi \times 31 {\rm{GHz}}, \nonumber \\
\omega_{B} &=& 2\pi \times 36 \rm{GHz}, \; \frac{\Omega_{A}}{V_{0}} \simeq \frac{\Omega_{B}}{V_{0}} = 2 \pi \times 2.3 \rm{\frac{GHz}{mV}}, \nonumber \\
D^{z} &=& 2 \pi \times 1.0 \rm{GHz}, \nonumber \\
D^{\pm} &=& 2\pi \times 1.4 {\rm{GHz}}.
\end{eqnarray}
A plot of $J$ for $V_{0} = 0.25,0.5 \rm{mV}$ is shown in figure fig.~\ref{Jvsphifig}, calculated from (\ref{Hlat}). For small values of $V_{0}$, $\abs{J}$ is almost completely independent of $\varphi_{s}$, but for larger $V_{0}$ the magnitude fluctuations become significant. $\abs{J}$ can be further increased by up to an order of magnitude by choosing device parameters to work in the regime where $f > 1/2$, but the relative qubit nonlinearities are smaller and the system becomes more susceptible to fluctuations in the external magnetic field. I emphasize that the parameters listed above certainly do not represent the best possible choice for many-body physics, and indeed, it may ultimately turn out that other types of qubits may be superior for reaching the bosonic fractional quantum Hall regime described below. Nonetheless, they demonstrate that my system could be engineered with current technology, and achieves hopping matrix elements which are around three orders of magnitude larger than the typical flux qubit decay and dephasing rates (around a MHz).

\subsection{Transmon Qubits}\label{transmon}

\begin{figure}
\includegraphics[width=3.5in]{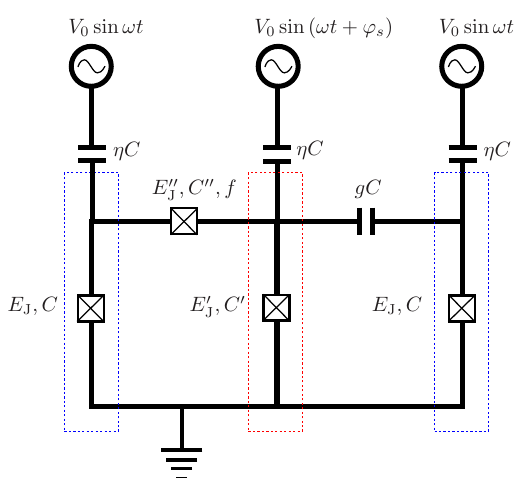}
\caption{(Color online) Implementation of the $\pm-B-zz$ link with transmon qubits. The effective circuit of a transmon qubit is a single Josephson junction connecting a small superconducting island to ground; the superconducting phase $\phi$ of the island is the qubit's single quantum degree of freedom. The $\pm$ coupling can be implemented by a capacitive coupling between two qubits ($gC$ in the figure). The $zz$ coupling is somewhat more challenging, but can be realized by connecting two qubits with a Josephson junction ($E_{\rm J}''$ and $C''$) with a flux bias $f$. The Josephson junction naturally produces $xx$, $yy$ and $zz$ couplings, and since both $xx$ and $yy$ couplings become $\pm$ couplings in the rotating frame, by tuning the flux bias $f$ and appropriately choosing $E_{\rm J}''$ and $C''$, we can get them to cancel each other out, leaving only the $zz$ term.}\label{transmonfig}
\end{figure}

I will now briefly outline the implementation of my model in transmon qubits. The basic circuit of a transmon qubit reduces to that of a single Josephson junction (with large $E_{\rm J}/E_{\rm C}$) connecting a small superconducting island to ground. The superconducting phase $\phi$ of the island is the qubit's sole quantum degree of freedom, and partly as a result of this, these qubits are extremely stable, achieving decay and dephasing times over an order of magnitude greater than flux qubits \cite{paikschuster,changvissers}. Further, requiring only a single Josephson junction, they are simpler to fabricate than flux qubits. This stability and simplicity comes at a cost, however, in that the natural nonlinearities of transmon qubits are only a few percent of the excitation energy $\omega$, which strongly constrains the magnitudes of the coupling terms.

The Transmon qubit Hamiltonian is a quantum anharmonic oscillator,
\begin{eqnarray}\label{HT}
H_{T} = -4 E_{\rm C} \frac{\partial^{2}}{\partial \phi^{2}} - E_{\rm J} \cos \phi.
\end{eqnarray}
The charge and phase operators map to $\sigma$ matrices in the qubit basis as they did in flux qubits:
\begin{eqnarray}\label{RT}
Q &\to & -2 e \MM_{\partial_{\phi}}^{01} \sigma^{y}, \; \sin \phi \to \MM_{\sin \phi}^{01} \sigma^{x},   \\
\cos \phi & \to & \frac{\MM_{\cos \phi}^{11}+\MM_{\cos \phi}^{00}}{2} \mathbf{1} + \frac{\MM_{\cos \phi}^{11}-\MM_{\cos \phi}^{00}}{2} \sigma^{z}. \nonumber
\end{eqnarray}
To construct the alternating $\pm$ and $zz$ couplings required to obtain tunable phases, we introduce alternating capacitive and flux-biased Josephson junction couplings. The capacitive part is simply a $\sigma^{y} \sigma^{y}$ term:
\begin{eqnarray}
H_{C} = +\frac{4 e^{2} g C''}{C C'}  \MM_{\partial_{\phi,A}}^{01} \MM_{\partial_{\phi,B}}^{10}  \sigma_{A}^{y} \sigma_{B}^{y}.
\end{eqnarray}
On the other hand, the Josephson junction with a flux bias (fig.~\ref{transmonfig}) contains $xx$, $yy$ and $zz$ terms by default; dropping $xz$ terms that will vanish in the rotating frame,
\begin{eqnarray}\label{HJJF}
H_{JJ} &=& + \frac{4 e^{2} C''}{C C'}  \MM_{\partial_{\phi,A}}^{01} \MM_{\partial_{\phi,B}}^{01}  \sigma_{A}^{y} \sigma_{B}^{y} \\
& &- \of{\cos^{2} 2\pi f - \sin^{2} 2\pi f} E_{\rm J}'' \times \nonumber \\
& & \left[ \MM_{\sin \phi,A}^{01} \MM_{\sin \phi,B}^{10} \sigma_{A}^{x} \sigma_{B}^{x} \right. \nonumber \\
& &  +2 \of{\MM_{\cos \phi,A}^{11}-\MM_{\cos \phi,A}^{00}} \nonumber \\
 & & \left. \times \of{\MM_{\cos \phi,B}^{11}-\MM_{\cos \phi_1,B}^{00}} \sigma_{A}^{z} \sigma_{B}^{z} \right]. \nonumber
\end{eqnarray}
Upon transitioning to the rotating frame, both $xx$ and $yy$ become $\pm$ couplings, so by tuning $f$, $E_{J}''$ and $C''$, we can cause the $\pm$ components to exactly cancel each other, leaving a pure $zz$ coupling. For appropriate $E_{J}''$ and $C''$, the bias field can be set to zero. The low-energy many-body Hamiltonian (\ref{Hlat}) will be the same whether the circuit is comprised of flux qubits, transmons, or a mix of the two, though the magnitudes of $J$ and $\Omega$ will of course vary from one implementation to the next. 

\subsection{An Alternative Formulation: Near-Resonantly Driven Transitions Between Excited States}\label{12drive}

An alternate, and potentially more attractive, formulation of this architecture in physical qubits is to exploit the nonlinearities of the physical device spectra to drive a transition between different excited states rather than between $\ket{0}$ and $\ket{1}$. Depending on the structure of these nonlinearities, such a drive signal may not be useful or possible, but for systems whose nonlinearities are very large and positive (flux qubits) or small and negative (transmons), driving the system at the $\ket{1} \leftrightarrow \ket{2}$ or $\ket{2} \leftrightarrow \ket{3}$ transition can have a number of practical advantages over the $\ket{0} \leftrightarrow \ket{1}$ formalism described in the previous section. In this subsection, I will demonstrate how artificial gauge fields can be generated in a system driven at $\omega_{12}$ instead of $\omega_{01}$, and work out the details of implementing this architecture in transmons or flux qubits.

Consider a superconducting qubit device whose low-lying states have energies $E_0 = 0, E_1 = \omega_{01}$ and $E_{2} = 2 \omega_{01} + \delta$, and let us assume that the energies higher excited states are all far detuned in energy from $\omega_{01}$ or $\omega_{01}+\delta$, so we can restrict ourselves to the basis of $\ket{0}$, $\ket{1}$ and $\ket{2}$. We now consider the familiar set of charge and phase operators $Q$, $\cos \phi$ and $\sin \phi$, and make the crucial assumption that the wavefunction describing state $\ket{1}$ in the phase basis has opposite parity in $\phi$ compared to $\ket{0}$ and $\ket{2}$. This property holds for both transmons and flux qubits operated at the $f=1/2$ symmetry point. Within the basis of $\ket{0}, \ket{1}$ and $\ket{2}$, we can write the charge and phase operators as: 
\begin{eqnarray}\label{ops012}
Q &=& \begin{pmatrix}
0 & i q_{12} & 0 \\
-i q_{12} & 0 & i q_{01} \\
0 & -i q_{01} & 0 
\end{pmatrix}, \\ 
\sin \phi &=& \begin{pmatrix}
0 &  s_{12} & 0 \\
 s_{12} & 0 & s_{01} \\
0 & s_{01} & 0 
\end{pmatrix},  \; \cos \phi  =  \begin{pmatrix}
c_{22} &  0 & c_{02} \\
 0 & c_{11} & 0 \\
c_{02} & 0 & c_{00} 
\end{pmatrix}. \nonumber
\end{eqnarray}
Here, $q_{ij} = \bra{i} Q \ket{j}$ and the other coefficients are defined analogously. Let us now consider the case of a single transmon qubit, where the nonlinearity $\delta$ is small and negative, driven by the voltage $V_0 \sin \of{\of{\omega_{01} + \delta + \epsilon} t + \varphi_{s} }$. In the rotating frame, the Hamiltonian for this qubit is:
\begin{eqnarray}\label{HTD}
H_{B \of{T}} = \begin{pmatrix}
- \delta - 2 \epsilon & \Omega_{12} e^{ i \varphi_s} & 0 \\
\Omega_{12} e^{ -i \varphi_{s}} & -\delta - \epsilon & \Omega_{01} e^{ i \varphi_s} \\
0 & \Omega_{01} e^{ -i \varphi_{s} } & 0
\end{pmatrix}
\end{eqnarray} 
We now choose parameters so that $\abs{\delta + \epsilon} \gg \Omega_{01}$. In this limit, the rotating frame eigenstates and energies of (\ref{HTD}) are:
\begin{eqnarray}\label{HTDstates}
 \cos \theta &=& \frac{\epsilon}{\sqrt{\epsilon^{2} + \Omega_{12}^{2} }}; \; \ket{0_B } = \cuof{0,0,1}, \\
 \ket{1_B} &=& \frac{ \cuof{ e^{2 i \varphi_s } \of{\cos \theta - 1}, e^{i \varphi_s} \sin \theta, 0} }{\sqrt{2 - 2 \cos \theta}}  \nonumber \\
 \ket{2_B} &=& \frac{ \cuof{ e^{2 i \varphi_s } \of{\cos \theta + 1}, e^{i \varphi_s} \sin \theta, 0} }{\sqrt{2 + 2 \cos \theta}} \nonumber \\
E_{0} &=& 0, \; E_{1/2} = -\delta - \epsilon \of{-/+} \sqrt{\epsilon^{2} + \Omega_{12}^{2} }. \nonumber
\end{eqnarray}
Since $\delta$ is negative, $\ket{0}$ is the ground state, and at resonance, $\theta = \pi/2$.

Having defined the single-qubit Hamiltonian above, we can now consider the coupling between two qubits $A$ and $B$, where as before $\varphi_s = 0$ for qubit $A$. As in the previous section, we can evaluate the hopping matrix elements from the three possible couplings $Q_A Q_B$, $\sin \phi_A \sin \phi_B$ and $\cos \phi_A \cos \phi_B$ to be
\begin{eqnarray}\label{HTDops}
\bra{0_A 1_B} Q_A Q_B \ket{1_A 0_B} = q_{01}^{2} \frac{\sin \theta}{\sqrt{2-2\cos \theta } } e^{-i \varphi_s }, \quad \\
\bra{0_A 1_B} \sin \phi_A \sin \phi_B \ket{1_A 0_B} = s_{01}^{2} \frac{\sin \theta}{\sqrt{2-2\cos \theta } } e^{-i \varphi_s }, \nonumber \\
\bra{0_A 1_B} \cos \phi_A \cos \phi_B \ket{1_A 0_B} = -c_{02}^{2} \sqrt{\frac{1-\cos \theta}{2}} e^{-2 i \varphi_s}. \nonumber
\end{eqnarray}
We readily see from these equations that a chain of ``$\pm$" and ``$zz$" couplings will again produce hopping matrix elements with arbitrarily tunable complex phases. The physical origin for these phases is as follows. In the rotating frame, the phase offset $\varphi_s$ causes the phase of physical qubit wavefunctions to advance by $\pm \varphi_s$ as they absorb or emit photons into the drive field, which can only induce transitions between $\ket{0} \leftrightarrow \ket{1}$ and $\ket{1} \leftrightarrow \ket{2}$. Since the first excited state in the rotating frame is a superposition of states $\ket{1}$ and $\ket{2}$ in the rest frame, a transition between it and the ground state driven by $Q$ or $\sin \phi$ will have a phase shift of $\pm \varphi_s$, since this process changes the state just as it would be changed by a single photon. The $\cos \phi$ operator, however, can mix states $\ket{0}$ and $\ket{2}$ in the rest frame, and thus acts as an effective two-photon process in a single step, advancing the phase by $\pm 2 \varphi_s$. Consequently, the $Q/\sin \phi$ ($\pm$) and $\cos \phi$ ($zz$) operators see the phase shifts differently, and the phase accumulated around a loop which chains together $\pm$ and $zz$ operators can be nonzero, indicating the presence of an artificial gauge field.

This method can be implemented identically in flux qubits, where $\delta$ is positive, and is between 6 and 24 times $\omega_{01}$ for $E_J = 40 E_C$ and $0.75 < \alpha < 0.85$. In this case, since the energy scales are so widely separated we can simply assume that a $\ket{0} \leftrightarrow \ket{1}$ transition from the drive field is forbidden, giving us the rotating frame Hamiltonian:
\begin{eqnarray}\label{HFQD}
H_{B \of{FQ}} = \begin{pmatrix}
\omega_{01} - \epsilon & \Omega_{12} e^{ i \varphi_s} & 0 \\
\Omega_{12} e^{ -i \varphi_{s}} & \omega_{01} & 0 \\
0 & 0 & 0
\end{pmatrix}.
\end{eqnarray} 
Aside from additional sign flips in the $Q$ and $\sin \phi$ operators which depend on which region of the qubit is being coupled to, the calculation proceeds identically as it did in transmons. In both cases, the physical wiring depicted in figs.~\ref{qubitfig} and \ref{transmonfig} can be left unchanged. 

Driving the qubits at this transition instead of $\ket{0} \leftrightarrow \ket{1}$ has a number of advantages. First, the energies of the rotating frame excited states are larger, avoiding the need for strong Rabi frequencies to get high energy excitations. Second, the hopping phase between two qubits can be tuned arbitrarily without changing the magnitude even if both qubits are on resonance; in the $\ket{0} \leftrightarrow \ket{1}$ case one of the qubits had to be significantly detuned to obtain these phases. Note, however, that if one wishes to tune the phase between any two sites without changing the phases between any other sites, one must still include ``auxilliary" qubits to mediate the tunneling. Third and most importantly, decays in the rest frame (such as an energy loss process which sends $\ket{1} \rightarrow \ket{0}$) cannot spontaneously create rotating frame excitations from the ground state when the qubits are driven near $\omega_{12}$, since the rotating and rest frame ground states are the same and do not mix with state $\ket{1}$. This means that the qubit array will be empty of excitations unless the system is populated by a second pulse near the rotating frame energy, making the population easier to control and eliminating a significant heating source in the many-body system. 

We see from these calculations that both flux qubits and transmons could be used to simulate artifcial gauge fields in exotic many-body systems. I will now describe both the simplest and most interesting of these systems: strongly interacting bosons in a uniform magnetic field, which would realize a bosonic fractional quantum Hall effect.

\section{Many-body States and the Lowest Landau Level}\label{LLL}

By considering a 2d lattice of qubits we arrive at the final hopping Hamiltonian (\ref{Hlat}). Previous studies \cite{hofstadter,kohmoto,assaad,hafezi,palmerkleinjaksch,sorensen,zhang,oktel,kapitmueller,onur,hormozimoller,kapitbraiding} have shown that the square lattice version of this Hamiltonian is analogous to the 2d lowest Landau level problem of strongly interacting bosons, and realizes abelian and non-abelian fractional ground states at the appropriate fixed densities. I expect that small arrays should be sufficient to observe quantum Hall physics, since the magnetic length $l_{B} = 1/\sqrt{2 \pi \Psi}$ (where $\Psi$ is the gauge-invariant phase accumulated when a particle circulates around a plaquette) can be less than a lattice spacing \footnote{I calculate the magnetic length by analogy to the mapping to the lowest Landau level in \cite{kapitmueller}; the coefficient of the Gaussian in the Landau level wavefunctions sets $l_{B}$.}. Connections between flux qubits beyond nearest neighbors can reproduce the exact lowest Landau level of the continuum \cite{kapitmueller,kapitbraiding} and lead to more robust fractional quantum Hall states, but they may not be necessary to observe the Laughlin state at $\nu = 1/2$ \cite{hafezi}. Here we adopt the standard definition of the filling fraction $\nu$ as the ratio of particle to flux density. A wide range of other possible quantum spin-1/2 models with 2-body interactions, both with complex phases and without, could be studied in this device architecture; I find quantum Hall systems to be the most intriguing, due to the existence of abelian anyons at $\nu = 1/2$ and the existence (with tuning) of non-abelian anyons at $\nu = 1$ and 3/2 \cite{nayaksimon,kapitbraiding}, along with other exotic states at different filling fractions. The boson density could be controlled by using a second external field at frequency $\omega'$ near the rotating frame energy $E_{A}$ to populate the lattice; the $\omega'$ dependence of the system's response to this field could be used to measure the gaps of the many-body states.


The incoherent particle loss rates in my array from single qubit decay and dephasing effects should not be a significant obstacle to studying strongly correlated many-body states. Using values from the previous section and from the superconducting qubit literature \cite{yoshihara,clarkewilhelm}, a typical hopping parameter would be $J/\hbar = 1 \rm{GHz}$. The decay rate would be roughly given by the decay/dephasing rate of the qubits, which for flux qubits is of order $1 \mathrm{MHz}$. With a Landau band spacing of $\omega_{\rm LLL}  \simeq 3 J$ in a square lattice at $\Psi = 1/4$ quanta per plaquette, the relative correction to the Landau bandwidth from this process would be thus be insignificant, provided that the system is driven to balance the incoherent particle loss. I expect that this loss rate by itself will not prevent quantum Hall states from forming in our array \cite{hafezilukin}. Likewise, a small number of ``dead" sites (where a qubit is defective and cannot be excited) should also be relatively harmless-- the many-body wavefunction can eliminate these defects simply by nucleating a quasihole at each site. So long as the density of flux quanta is large compared to the defect density, these defects will simply make small shifts in the gap energy and particle density of the gapped states, but will not have any other qualitative effects on the system.

More worrisome is the issue of time-independent random variations in the qubit properties at every site, which could disrupt the formation of topological states if these variations became large enough. To quantify this issue, I numerically simulated the broadening of the lowest Landau level in our model as a function of three static (quenched) noise sources: random fluctuations in the on-site potential (shifts in the rotating frame excitation energy of a given $A$ qubit), random fluctuations in the magnitude of $J$, and random fluctuations in the phase of $J$ between neighboring sites. In a real system, these noise sources would be correlated, but as the details of those correlations would depend in part on the physical implementation of the qubits, I have assumed that each type of quenched disorder is applied randomly to every site with no dependence on the other types or on the disorder at nearby sites. To determine the broadening from each noise source, I numerically diagonalized the single-particle hopping matrix on 8$\times$8 and 12$\times$12 lattices with periodic boundary conditions, given by the Hamiltonian:
\begin{eqnarray}\label{deltaH}
H_{LLL} &=& \sum_{ij} \of{F_{ij}} J_{ij} \of{ e^{i \of{ \phi_{ij} + \pi \delta \phi_{ij}}} + H. C.} \\ & & + \sum_{i} J_{NN} \delta U_{i} n_{i}. \nonumber
\end{eqnarray}
Here, the hopping matrix elements are restricted to nearest and next nearest neighbors with relative magnitudes chosen as in \cite{kapitmueller}, $\delta U_{i}$ and $\delta \phi_{ij}$ are dimensionless parameters which are Gaussian distributed about 0, $J_{NN}$ is the average nearest neighbor hopping energy, and $F_{ij}$ is a dimensionless parameter which is Gaussian distributed about 1. I diagonalized (\ref{deltaH}) for 25 random distributions of noise for each data point (in steps of 0.02 for each $\sigma$), and from the spectrum I extracted the lowest Landau level broadening $\Delta$, which is the ratio of the energy splitting between the lowest and highest LLL states to the splitting between the highest LLL state and the bottom of the first excited band. I then fit $\Delta \of{\sigma_{U/J/\Psi}}$ as a function of the standard deviation of each noise source with the other two sources set to zero; this relationship was linear in each case for small fluctuations. The results of our simulations are shown in table~\ref{LLLtable}; note that $\Delta_0$ is nonzero even without defects, as a consequence of truncating the Hamiltonian in \cite{kapitmueller} to nearest and next nearest neighbor hopping.

It is important to note that this calculation only captures distortions to the single particle spectrum and that the many-body response to noise of this type is a subtle problem beyond the scope of this work. However, one should qualitatively expect that the topological states should be disrupted when the normalized Landau level splitting $\Delta$ approaches the dimensionless quasiparticle excitation gap $\Delta_{qp}/J_{NN}$. In numerical studies of this system in the clean limit with hard-core 2-body repulsion (largely unpublished), $\Delta_{qp}/J_{NN}$ typically ranged between $0.2$ and $1$ for correlated states at different flux and particle densities, and tended to be larger at smaller filling fractions. This suggests that many-body quantum Hall states should exist in my system when noise is sufficiently well-controlled.

\begin{table}
\begin{tabular}{| c | c | c | c | c |}
\hline
Flux Density & $\Delta_{0}$ & $C_{U}$ & $C_{J}$ & $C_{\Psi}$ \\
1/4 & 0.015 & 0.41 & 1.42 & 1.75 \\
1/3 & 0.018 & 0.72 & 1.21 & 2.36 \\
3/8 & 0.08 & 0.35 & 0.99 & 1.94 \\
\hline
\end{tabular}
\caption{Robustness of the lowest Landau level to external noise sources. For the random noise simulations described in the text, I fit the normalized splitting $\Delta$ of the lowest Landau level to the function $\Delta = \Delta_{0} + C_{U} \sigma_{U} + C_{J} \sigma_{J} + C_{\Psi} \sigma_{\Psi}$, where the $\sigma$'s are the standard deviation of each noise source (local potential, hopping magnitude, and hopping phase) which is applied randomly to every site ($\sigma_{U}$) and link between sites ($\sigma_{J}$ and $\sigma_{\Psi}$). As seen in the Hamiltonian (\ref{deltaH}), the potential fluctuations are in units of $J_{NN}$ and the phase fluctuations are in units of $\pi$. Above the flux density $\Psi = 1/3$, truncation to nearest and next nearest neighbor hopping introduces significant broadening even in the clean system, so flux densities of 1/3 or less should be the focus of experiments on our design.}\label{LLLtable}
\end{table}


\section{A Simple Experiment to Demonstrate the Gauge Field}\label{ex}

\begin{figure}
\includegraphics[width=3.25in]{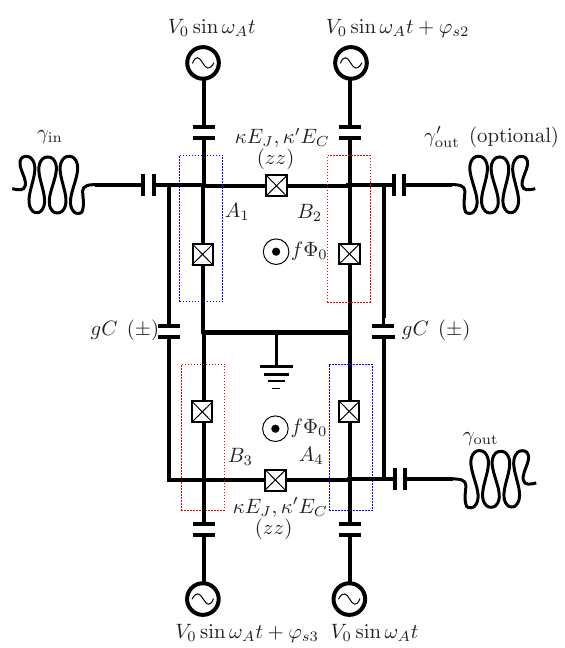}
\caption{(Color online) Configuration to demonstrate the artificial gauge field in four transmon qubits (blue and red boxed regions), as outlined in section~\ref{ex}. As described in section~\ref{ex}, appropriately tuning the phase offsets $\varphi_{si}$ will produce a gauge-invariant phase difference in the two paths that the mobile fluxon excitation could take from qubit 1 to qubit 4. The resulting interference of these two paths can be detected in the time-dependent probability $P_{4} \of{t}$ of qubit 4 being in its excited state. Additional flux biases convert the single Josephson junction couplings between $A$ and $B$ transmon qubits to pure $zz$ interactions.}\label{4Qfig}
\end{figure}

While the ultimate purpose of this proposal is to study exotic many-body states in an array of hundreds or thousands of flux qubits, the existence of a nontrivial gauge field can be demonstrated by studying an arrangement of four flux qubits, connected in a loop. Consider a square loop of four flux qubits labeled (1-4), where qubit 1 sits at the top left corner and qubit 4 at the bottom right, as shown in Fig.~\ref{4Qfig}. For this choice, any hop through a $D^{z}$ coupling will accumulate a phase $\psi$, giving a total of $\Psi = 2 \psi$ for a complete circuit of the loop.  Conversely, if the phases of the voltages applied to the $B$ qubits are shifted by $\pi$ from one qubit pair to the next, the magnitude of the hopping matrix element will be unchanged but there will be no complex phase accumulation. In this case, the $B$ qubits have identical rotating frame energies to the $A$ qubits, and differ from them through the relative phases $\varphi_{si}$ of the applied voltages. We will assume for simplicity that the magnitudes of the hopping matrix elements from the $D^{z}$ and $D^{\pm}$ couplings are both equal to $J$.

To demonstrate that the alternating voltages generate a nonzero effective flux through the four-qubit loop, we first initialize the array by letting all four qubits relax to their ground states. At time $t=0$, we apply a microwave pulse to qubit 1 to excite it into the rotating frame excited state $\ket{1_A}$, and then at time $t$ we measure the state of qubit 4. The probability of qubit 4 being excited is given by
\begin{eqnarray}\label{P4}
P_{4} \of{t} &=& \abs{ \bra{0_{1} 0_{2} 0_{3} 1_{4} } e^{i H t/\hbar} \ket{1_{1} 0_{2} 0_{3} 0_{4} } }^{2} \\
&=& \frac{1}{4} \of{ \cos \of{ \frac{2 t J}{\hbar} \cos \frac{\Psi}{4} } - \cos \of{ \frac{2 t J}{\hbar} \sin \frac{\Psi}{4} } }^{2}. \nonumber
\end{eqnarray} 
This interference pattern is particularly striking when $\Psi$ is nearly equal to $\pi$. If we let $\Psi = \pi + \epsilon$, the probability distribution becomes
\begin{eqnarray}
P_{4} \of{t} = \of{ \sin \frac{\sqrt{2} J t}{\hbar} }^{2} \of{ \sin \frac{J t \epsilon}{2 \sqrt{2} \hbar} }^{2}.
\end{eqnarray}
In the limit of $\epsilon \to 0$, the probability of qubit 4 being occupied becomes zero at all times, due to the perfect interference of the two paths. This is a dramatic effect, and while field fluctuations and fabrication defects would prevent perfect interference in a real device, the strong slowing of the occupation periodicity of qubit 4 as $\Psi$ approaches $\pi$ would be readily observable. Such interference is only possible if there is a gauge-invariant phase difference between the two paths, and would therefore demonstrate that nontrivial effective gauge fields are realized in my architecture. 

An alternative experiment would be to connect a microwave source to qubit 1 and a microwave drain at qubit 2, and measure the transmission coefficient as a function of $\Psi$ for photons near the rotating frame excitation energy $E_A$. At $\Psi = 0$ the transmission coefficient should be maximal, and at $\Psi = \pi$ it should be zero (or nearly zero when defects are taken into account), owing to the destructive interference of the two paths which is the key signature of a gauge field. I note also that a similar arrangement of three qubits with two $\pm$ links and one $zz$ link could potentially engineer a charge noise free variant of the circulator design in \cite{kochhouck}; as a microwave circulator requires time reversal symmetry breaking to function, it could also demonstrate the existence of a nontrivial gauge field.

\section{Conclusion}

I have demonstrated a method for realizing a quantum Hall state of bosons using asymmetric qubit pairs, driven by applied oscillating electric fields. I also demonstrated that my model could be implemented in lattices of flux or transmon qubits. With appropriate protocols for stabilizing the average particle density and measuring the conductivity, I expect that conductivity quantization could be observed on small arrays, though I note that the details of how to measure the conductivity are beyond the scope of this article. The statistics of anyonic collective modes could be determined through similar methods \cite{dassarma,bondersonkitaev,willett,anbraiding}. 

Further, the dynamical tunability of my model could be exploited to realize exotic combinations of states that would be difficult or impossible to study in cold atom or solid state systems. One could locally adjust the applied voltage $V_{0} \sin \of{\omega t + \varphi_s}$ and flux bias $f$ to change the gauge field density and effective chemical potential in a given region, creating islands of arbitrary shape which could be at a different filling fraction than the surrounding lattice and thus have different anyonic modes. Alternately, by reversing the signs of all the phase shifts $\varphi_s$ in a region, one can crate a sharp boundary between regions with effective gauge fields of equal magnitude but opposite sign. In both cases, we expect physics along the boundaries to be rich.

Finally, by locally tuning $V_{0}$, $\varphi_s$ and $f$ to manipulate vortices in the qubit lattice, arrays of ordinary qubits could be used to construct a topological non-abelian anyon qubit \cite{kitaev2003,hormozi,nayaksimon}, trading information density for topological protection against decoherence. Though far down the road, in that sense my proposal could be similar to the surface code and cluster state \cite{yaowang,fowlersurface} ideas developed in recent years, and could provide a new potential mechanism for reducing decoherence in superconducting quantum information devices.

\section{Acknowledgments}

I would like to thank John Chalker, Greg Fuchs, Chris Henley, Matteo Mariantoni, John Martinis and David Pappas for useful discussions related to this project. 
I am indebted to Paul Ginsparg for his critical comments and advice, and to Steve Simon for his assistance in the final preparation of this manuscript. Most of all, I would like to thank Erich Mueller for his guidance at many stages of this project. This material is based on work supported in part by the National Science Foundation Graduate Program, EPSRC Grant No. EP/I032487/1, and Oxford University.



\bibliography{FQgauge}

\end{document}